\documentclass[apj]{emulateapj}
\usepackage{amsmath,amssymb,graphicx,color,longtable,natbib}

\begin{document}
\normalsize

\slugcomment{Accepted for publication in The Astrophysical Journal}

\title{GERLUMPH Data Release 1: High-resolution cosmological microlensing magnification maps and eResearch tools}

\author{G.~Vernardos\altaffilmark{1},  C.J. Fluke\altaffilmark{1} N.F. Bate\altaffilmark{2} and D. Croton\altaffilmark{1}}
\affil{Centre for Astrophysics \& Supercomputing, Swinburne University of Technology, PO Box 218, Hawthorn, Victoria, 3122, Australia}
\affil{Sydney Institute for Astronomy, School of Physics, A28, University of Sydney, NSW, 2006, Australia}

\begin{abstract}
As synoptic all-sky surveys begin to discover new multiply lensed quasars, the flow of data will enable statistical cosmological microlensing studies of sufficient size to constrain quasar accretion disc and supermassive black hole properties.
In preparation for this new era, we are undertaking the GPU-Enabled, High Resolution cosmological MicroLensing parameter survey (GERLUMPH).
We present here the GERLUMPH Data Release 1, which consists of 12342 high resolution cosmological microlensing magnification maps and provides the first uniform coverage of the convergence, shear and smooth matter fraction parameter space.
We use these maps to perform a comprehensive numerical investigation of the mass-sheet degeneracy, finding excellent agreement with its predictions.
We study the effect of smooth matter on microlensing induced magnification fluctuations.
In particular, in the minima and saddle-point regions, fluctuations are enhanced only along the critical line, while in the maxima region they are always enhanced for high smooth matter fractions ($\approx 0.9$).
We describe our approach to data management, including the use of an SQL database with a Web interface for data access and online analysis, obviating the need for individuals to download large volumes of data.
In combination with existing observational databases and online applications, the GERLUMPH archive represents a fundamental component of a new microlensing eResearch cloud.
Our maps and tools are publicly available at {\tt http://gerlumph.swin.edu.au/}.
\end{abstract}

\keywords{gravitational lensing: micro -- accretion, accretion discs -- quasars: general}

\section{Introduction}
\label{sec:intro}
Gravitational lensing studies the effect of matter on the propagation of light through the universe.
Observationally, gravitational lensing is characterised by the creation of multiple images or arcs (strong lensing), coherent shape distortions (weak lensing), or high magnifications due to compact objects \citep[microlensing;][]{Schneider2006}.
Quasar microlensing, the effect on the magnification of individual images in a multiply-imaged quasar due to an ensemble of stellar-mass compact objects near the line of sight, is now well established as a tool for studying the structure of quasars \citep{Schmidt2010}.

The magnitude of the microlensing effect depends strongly on the size of the emitting source; smaller sources produce more significant microlensing induced magnification variations \citep{Wambsganss1990b}.
The relevant scale here is the Einstein radius, $R_{\rm Ein}$, the radius of the symmetric ring that occurs when a source is directly aligned with a gravitational lens or microlens:
\begin{equation}
R_{\rm Ein} = \sqrt{ \frac{D_{\rm os}D_{\rm ls}}{D_{\rm ol}} \frac{4G\langle M \rangle}{c^2} } \, .
\end{equation}
This depends on the angular diameter distances from observer to lens, $D_{\rm ol}$, observer to source, $D_{\rm os}$, and lens to source, $D_{\rm ls}$, and the mean mass of the microlenses $\langle M \rangle$.
A typical value for $R_{\rm Ein}$ is $5.35 \pm 1.2 \times 10^{16}$ cm, which is the mean of 59 lensed systems from the CASTLES\footnote{http://www.cfa.harvard.edu/castles/} project that have both redshifts for the lens and source available (for $\langle M \rangle = 1$ M$_{\odot}$ and $H_{\rm 0} = 70$ km s$^{-1}$ Mpc$^{-1}$).

Since its suggestion by \cite{Chang1979} and subsequent discovery by \cite{Irwin1989}, microlensing has been used to study the physical and geometrical properties of quasars, ranging from the broad emission line region (BELR; $\sim10^{17}$cm), to the accretion disc ($\sim10^{15}$cm) surrounding the central supermassive black hole.
For a few systems, the size and temperature profiles of accretion discs have been found to be generally consistent with the \cite{Shakura1973} thin-disc model \citep{Anguita2008,Bate2008,Eigenbrod2008,Mosquera2011a,Blackburne2011}.
However, \cite{Floyd2009} rule out this model for the gravitational lens system SDSSJ0924+0219.
\cite{Morgan2010} use microlensing deduced properties of the disc to study the central supermassive black hole for 11 quasars, and although the results agree with a thin disc model, a very low radiative efficiency ($\sim$0.01\%) is implied.
\cite{Dai2010} found that the X-ray emission of RXJ1131-1231 is located in a small region ($\sim10^{14}$cm) near the central black hole.
\cite{Sluse2012} studied 17 lensed quasars and found that the geometry of the BELR is not necessarily spherically symmetric.

The above studies have focused either on single objects or small samples from the $\sim90$ currently known lensed quasars \citep{Mosquera2011b}.
This is because long-term monitoring and/or multiwavelength observations require a lot of effort and resources, e.g. \cite{Poindexter2008} use data spanning a 13-year period and 11 bands for the double lens HE1104-1805.
However, this is about to change with the new generation of synoptic all-sky survey telescopes, such as the Large Synoptic Survey Telescope \citep[LSST; ][]{LSST2009}, Pan-STARRS \citep{Kaiser2002}, and SkyMapper \citep{Keller2007}, which are expected to survey the entire sky regularly.
These facilities are expected to discover a few thousand microlensing candidates \citep{Oguri2010}, and provide regular monitoring without any additional effort.
As the number of lensed quasars increases, it becomes ever more important to move from single-object studies to statistically meaningful samples.

\begin{figure*}[t]
\includegraphics[width=\textwidth]{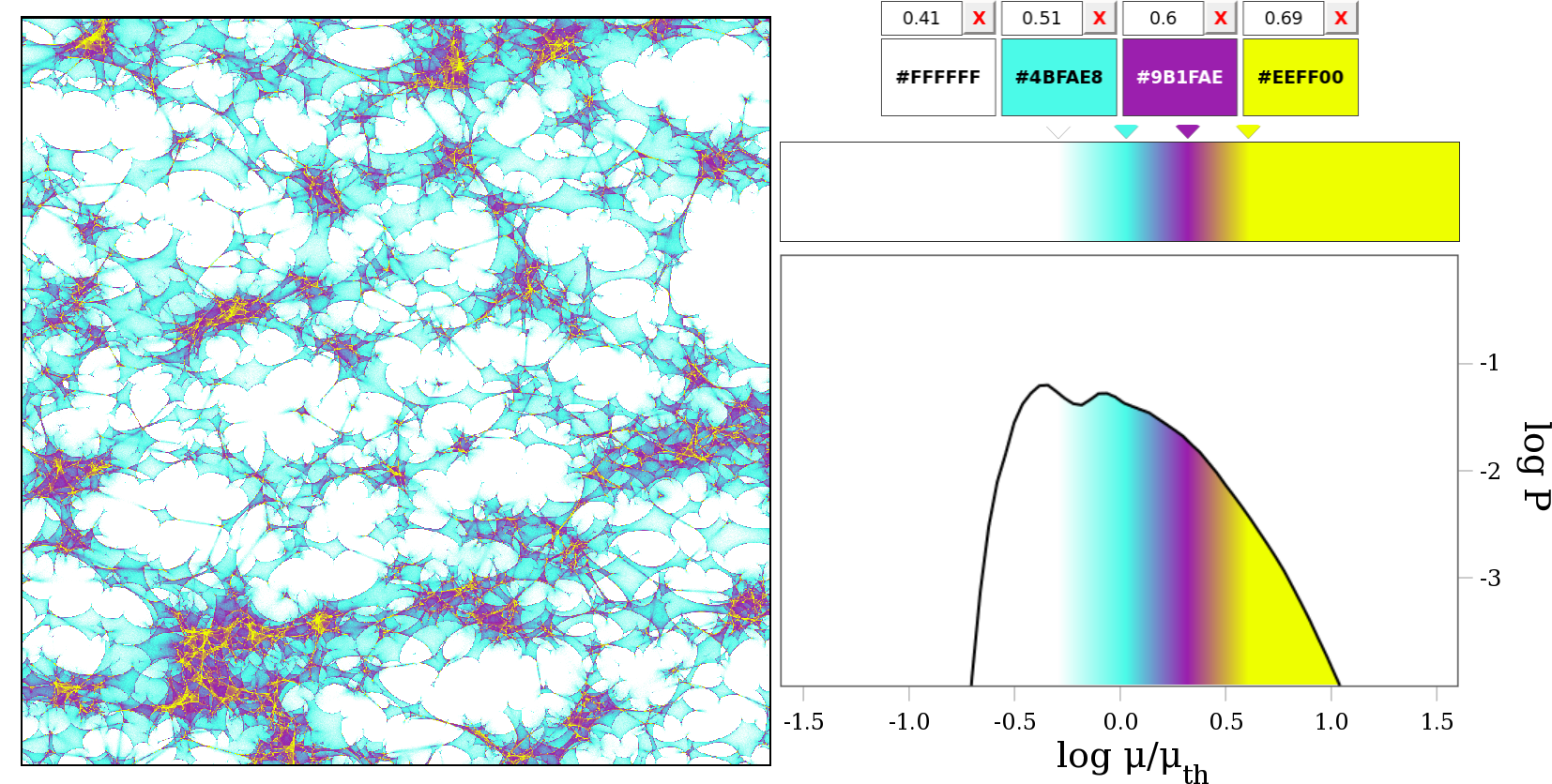}
\caption{An example of the usage of the `Colorbar' tool. The caustic network of a magnification map can be highlighted with different colors at the same time as its corresponding magnification probability distribution. In this example we can see areas of low magnification, or demagnification, ($\mu < 0.3 \mu_{\rm th}$) shown in white, areas of almost no magnification ($\mu \approx \mu_{\rm th}$) shown in cyan, areas of high magnification ($\mu \approx 3 \mu_{\rm th}$) shown in purple and finally areas of very high magnification ($\mu > 3 \mu_{\rm th}$) shown in yellow.\label{fig:colorbar}}
\end{figure*}

On the theoretical side, the basic tool for microlensing studies is the magnification map (hereafter `map'): a statistical representation of the combined effects of an ensemble of microlenses presented as a pixellated image in the source plane (a typical network of caustics i.e. regions of high magnification in the source plane, can be seen in the example map of Figure \ref{fig:colorbar}).

A map is defined in terms of the convergence, $\kappa$, which describes the combined focusing power of the compact microlenses, $\kappa_{*}$, and smooth matter, $\kappa_{\rm s}$, where $\kappa = \kappa_{*} + \kappa_{\rm s}$, and the shear, $\gamma$, which describes the distortion applied due to the external mass distribution of the lens galaxy.
A further useful parametrization is to define the smooth matter fraction, $s=\kappa_{\rm s}/\kappa$.
Additional parameters required to generate a magnification map, namely the mass and positions of the microlenses, and the width, resolution and statistical accuracy of the map, are discussed in Section \ref{sec:survey}.
For studies of specific systems, additional physical parameters may need to be introduced e.g. motions of the microlenses \citep{Mosquera2013}.

Magnification maps can be used to study the size and geometry of the source by extracting model light-curves \citep[the light-curve method, e.g.][]{Morgan2010}, or the temperature profile of the accretion disc by producing probability distributions of the flux in different wavelengths \citep[the snapshot method, e.g.][]{Bate2008}.
Producing a magnification map is a computationally demanding task, however, there are a few methods available \citep{Wambsganss1999,Kochanek2004,Thompson2010,Mediavilla2011b}, all of which use a variation of the inverse ray-shooting technique \citep{Kayser1986}.

In preparation for the imminent discoveries of new microlensed quasars by the future surveys, a systematic exploration of the microlensing parameter space should be considered \citep[hereafter BF12]{Bate2012}.
In this work, we report on the first results from the GPU-Enabled, High Resolution, cosmological MicroLensing parameter survey (GERLUMPH), using the $\sim$100 teraflop s$^{-1}$ GPU-Supercomputer for Theoretical Astrophysics Research (gSTAR) and the direct inverse ray-shooting method of \cite{Thompson2010} to address the time-consuming map calculations.

This paper describes the GERLUMPH Data release 1 (GD1) maps, which comprise a complete set of high resolution maps in the $\kappa,\gamma$ and $s$ parameter space.
Moreover, we also release the GERLUMPH Data 0 (GD0) maps, the dataset used in \citet[hereafer VF13]{Vernardos2013}.
The properties of the two datasets are summarized in Table \ref{tab:datasets} and presented in more detail in Section 2, where we describe our approach to a microlensing parameter survey and compare it to BF12.
In Section 3, we present the results of two initial applications: an extensive numerical investigation of the mass-sheet degeneracy, and a study of the effect of smooth matter on microlensing induced magnification fluctuations, throughout the $\kappa,\gamma$ and $s$ parameter space.
We make our data publicly available via a web server\footnote{http://gerlumph.swin.edu.au/} and provide online tools for map analysis, described in Section 4.
Discussions on our results and eResearch approach follow in Sections 5 and 6.
We present our conclusions in Section 7.

\section{Survey specifications}
\label{sec:survey}
We have used our GPU-D ray-shooting code (\citealt{Thompson2010,Bate2010}; VF13) in combination with the gSTAR supercomputer facility to generate the GD1 dataset of 12342 magnification maps.
The computation of the caustic structure in each map depends on eight parameters, which can be categorized into three groups: macromodel parameters, parameters of the microlenses and map characteristics.
Our choice of parameters for GD1 (and GD0) is shown in Table \ref{tab:datasets} and discussed below:

\begin{deluxetable}{rll}
\tabletypesize{\scriptsize}
\tablewidth{0pt}
\tablecaption{The GERLUMPH datasets\label{tab:datasets}}
\tablehead{\colhead{parameters} & \colhead{GD0} & \colhead{GD1}}
\startdata
$N_{\rm \kappa,\gamma}$         & 170          & 1122   \\
$N_{\rm s}$                     & 1            & 11    \\
Width ($R_{\rm Ein}$)           & 24           & 25    \\
Resolution (pixels)             & $4096^2$     & $10000^2$ \\
$N_{\rm sets}$                  & 15           & 1     \\
Total maps                      & 2550         & 12342 \\
GPU time (days)                 & 213          & 2902  \\
data size (TB)                  & 0.16         & 4.5   \\
\enddata
\tablecomments{$N$ denotes the number of different values for each parameter e.g. $N_{\rm \kappa,\gamma}$ is the number of individual $\kappa,\gamma$ grid locations, $N_{\rm sets}$ indicates the different sets of random microlens positions, etc. GPU time is the computational time used to generate the maps on a single GPU.}
\end{deluxetable}

\subsection{Macromodel (external) parameters}
\label{sec:macro}
In order to understand multiply imaged systems, one has to assume a model for the mass distribution of the galaxy-lens \citep[e.g.][]{Schneider2006}.
These models, hereafter `macromodels', are chosen on the basis of how well they can reproduce a number of observed properties e.g. the positions of the multiple (macro) images.
The values of $\kappa,\gamma$ are extracted from the macromodel at each position on the lens plane and can be used to calculate the magnification of each macroimage,
\begin{equation}
\mu_{\rm th} = \frac{1}{(1-\kappa)^2 - \gamma^2} \, .
\end{equation}
$\kappa,\gamma$ can then be used in the following stage of microlensing modelling i.e. generating magnification maps.

Our maps cover the $\kappa,\gamma$ parameter space uniformly: $0.0 < \kappa < 1.7$, $0.0 \leq \gamma \leq 1.7$,  with $\Delta \kappa, \Delta \gamma = 0.05$.
Smooth matter is taken into account for each $\kappa,\gamma$ combination by generating maps for $0 \leq s \leq 0.9$, with $\Delta s = 0.1$, and $s=0.99$.
The coverage of the $\kappa,\gamma$ parameter space by GD1 is shown in Figure \ref{fig:existing}, together with existing values for 23 multiply lensed systems, extracted from a number of macromodels from literature (BF12).
The critical line, i.e. where $\mu_{\rm th} \rightarrow \infty$, divides the $\kappa,\gamma$ parameter space into the minima, saddle-point and maxima regions.
These regions correspond to the extrema of the time-delay surface, which is where the macroimages form \citep[see][for details]{Blandford1986}.

It is important to point out here that there is a limit to how uniquely determined the macromodel derived $\kappa,\gamma$ values can be due to the mass-sheet degeneracy \citep{Falco1985,Gorenstein1988}.
Scaling the mass distribution of the lens and adding a homogeneous surface mass density (mass-sheet) will result in a transformation of coordinates in the source plane, which cannot be directly observed, leaving all other observables unchanged viz. image positions and shapes, flux ratios, etc.
This means that we cannot uniquely determine the lens-galaxy mass distribution, and consequently the resulting $\kappa,\gamma$, without additional information on the source (e.g. absolute size or luminosity) or on the lens-galaxy (e.g. mass derived from observations of stellar dynamics).

\subsection{Parameters of the microlenses}
It has been known that the mass of the microlenses has a negligible effect on the magnification probability distribution (MPD) of a magnification map over most of parameter space \citep{Wambsganss1992,Lewis1995,Wyithe2001,Schechter2004}.
Therefore, we adopt the simplest treatment of the microlens mass function viz. a constant mass of 1 M$_{\odot}$.

VF13 used GD0 to study systematic map properties for the case of compact matter only ($s=0$).
One of their results was on the effect of randomly positioning the microlenses on the lens plane.
It was found that changing the microlens positions leads to statistically equivalent maps over most of $\kappa,\gamma$ parameter space.
However, there are regions of parameter space i.e. the maxima region and along the critical line for $\kappa \lesssim 1$ (figure 6, VF13), where one particular set of microlens positions may lead to a quite different MPD (but still an equally valid choice).
Therefore, more than one map would be needed in those regions for subsequent calculations to be representative.
We have chosen to calculate a single map per $\kappa,\gamma,s$ combination, keeping in mind that our results may be affected by this known systematic.

\subsection{Map characteristics}
\label{sec:fluct_intro}
VF13 find that the smaller the width of a map, the more likely it is to get statistically different maps, for a given combination of $\kappa,\gamma$ (figure 7, VF13).
The GD1 maps have a width of 25 $R_{\rm Ein}$, which is high enough to minimize this effect, while still keeping the necessary computations within our capabilities.

The map resolution is set to $10000^2$ pixels.
This corresponds to 0.0025 $R_{\rm Ein}$, or $1.34 \times 10^{15}$ cm for a typical value of $R_{\rm Ein}$.
These maps are suitable for studying spatial scales from the accretion disc to the BELR for the sample of known multiply imaged quasars (see section 2.1 of BF12 for a justification of this choice).

Microlensing effectively produces deviations from the macromodel predicted magnification of a background source, towards high and low magnifications \citep{Paczynski1986}.
Both cases are of interest, the former for studying caustic crossing events \citep[e.g.][]{Witt1993,Anguita2008} and the latter in the case of anomalous flux systems \citep[e.g.][]{Schechter2004,Bate2008}.
The inverse ray-shooting technique measures the magnification by shooting a large number of rays through the lens plane and mapping them on the source plane.
Low magnification means that a small number of rays reached the examined position in the source plane.
Therefore, for a map to accurately probe the ranges of magnification of interest, we need to shoot a very large number of rays, $\mathcal{O}(10^{10})$.
The final average number of rays per map pixel, $N_{\rm avg}$, among the GD1 maps is $457\pm26$, which is high enough (the systematics of $N_{\rm avg}$ have been examined by VF13, who found an average of $302\pm24$ rays per pixel to be sufficient).

\subsection{GD1 data}
The total data size for GD1 is 4.5 Terabytes (TB).
Maps are stored in binary format, using unsigned integers to represent the number of rays per pixel, leading to a filesize of 381 MB per map.
BF12 report smaller filesizes and suggest using an unsigned short integer data type.
In our case however, we have ray counts that exceed 65,536, the maximum number for unsigned short integer representation, due to the much larger total number of rays that we are shooting.
gSTAR has $\sim1.7$PB of storage space, therefore storing a few tens of thousands of high resolution maps is not a problem.
Our web server configuration, and how to access and download the GD1 maps, is described in Section \ref{sec:server}.

Filesize may become important for users downloading tens or hundreds of maps from the GERLUMPH web server.
Longer term, our preferred model of operation for parameter-space investigations is fully remote analysis via online e-tools, obviating the need for individuals to download large volumes of data. For now, we use standard Unix compression tools ({\tt bzip2}) to compress the binary files delivered to the users over the internet, reducing file sizes by $\approx65\%$.

\begin{figure}[t]
\plotone{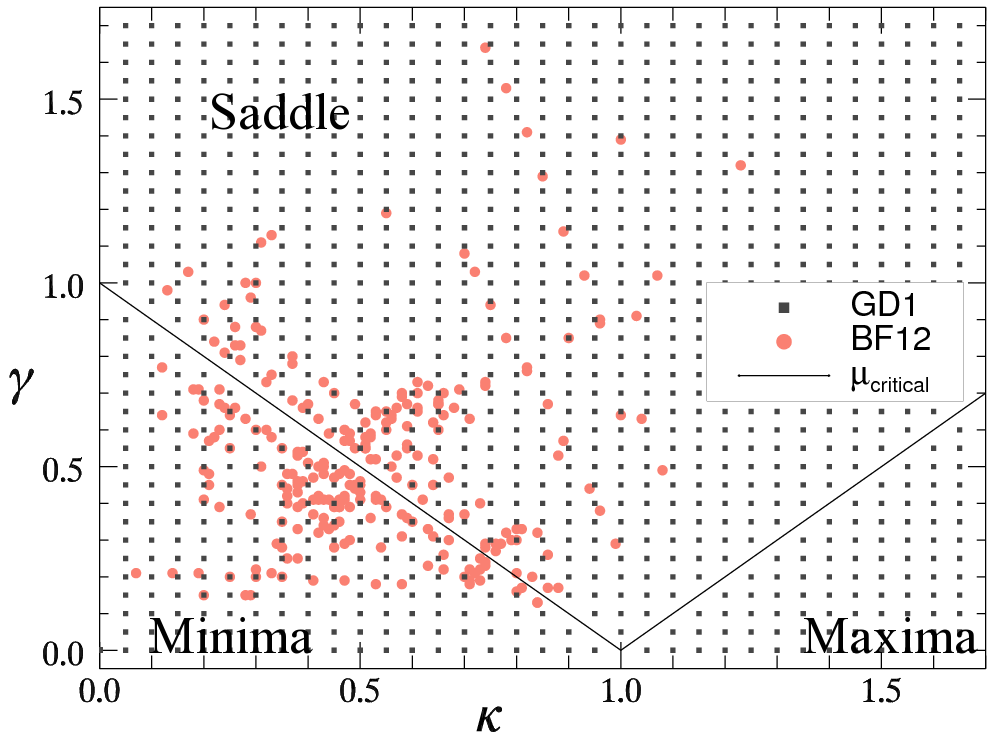}
\caption{$\kappa,\gamma$ values used in GD1 (black squares) and the 266 unique $\kappa,\gamma$ pairs from existing macro-models (circles, light red online) as compiled by BF12. GD1 maps cover the range of the existing macro-models; there is always a GD1 map within $\Delta \kappa,\Delta \gamma \leq 0.025$ of a model point. The solid line ($\mu_{\rm critical}$) is the critical line, i.e. the locus of $\mu_{\rm th} \rightarrow \infty$. An interactive version of this plot and the compilation of BF12 can be found online at the macro-model section of the GERLUMPH server.\label{fig:existing}}
\end{figure}

\section{Results}
\label{sec:results}
\begin{figure}[t]
\plotone{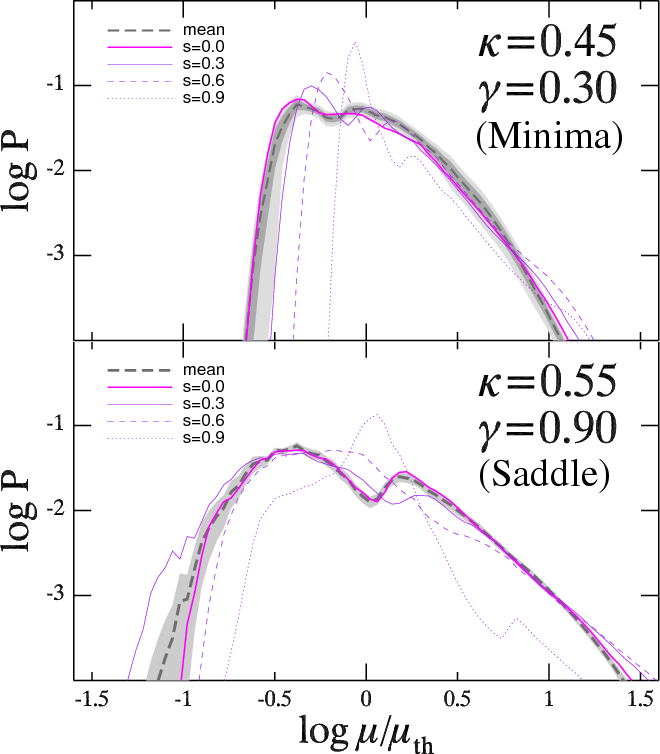}
\caption{The magnification probability distribution (MPD) of a GD1 $10000^2$-pixel map, thick solid line (magenta online), compared to the mean MPD obtained by VF13, thick dashed line, for $\kappa,\gamma,s$ equal to $(0.45,0.3,0)$, top panel, and $(0.55,0.9,0)$, bottom panel. The grey area is one standard deviation from the mean MPD. The thin solid, dashed and dotted lines (purple online) are the MPDs for $s$ equal to 0.3, 0.6 and 0.9 respectively. This Figure can be reproduced for any value of $\kappa,\gamma$ via the tools section of the GERLUMPH server.\label{fig:single}}
\end{figure}

We begin by examining the agreement between the MPDs of the GD1 maps and the mean MPD determined by VF13 from 15 different maps at lower resolution ($4096^2$ pixels) in the case without smooth matter.
As an example, in Figure \ref{fig:single} we compare the mean MPD, thick dashed line, and the MPD of GD1, thick solid line, for two trial combinations, $\kappa,\gamma=(0.45,0.3)$, corresponding to the minima region, and $\kappa,\gamma=(0.55,0.9)$, corresponding to the saddle-point region.
Apart from very low magnifications ($\mu < 0.1 \mu_{\rm th}$), the high resolution MPD agrees well with the lower resolution mean MPD.
The behaviour for low $\mu$ is expected due to the low ray counts (small number statistics), however, the higher resolution MPD is still within one standard deviation from the mean.

\begin{figure}[t]
\plotone{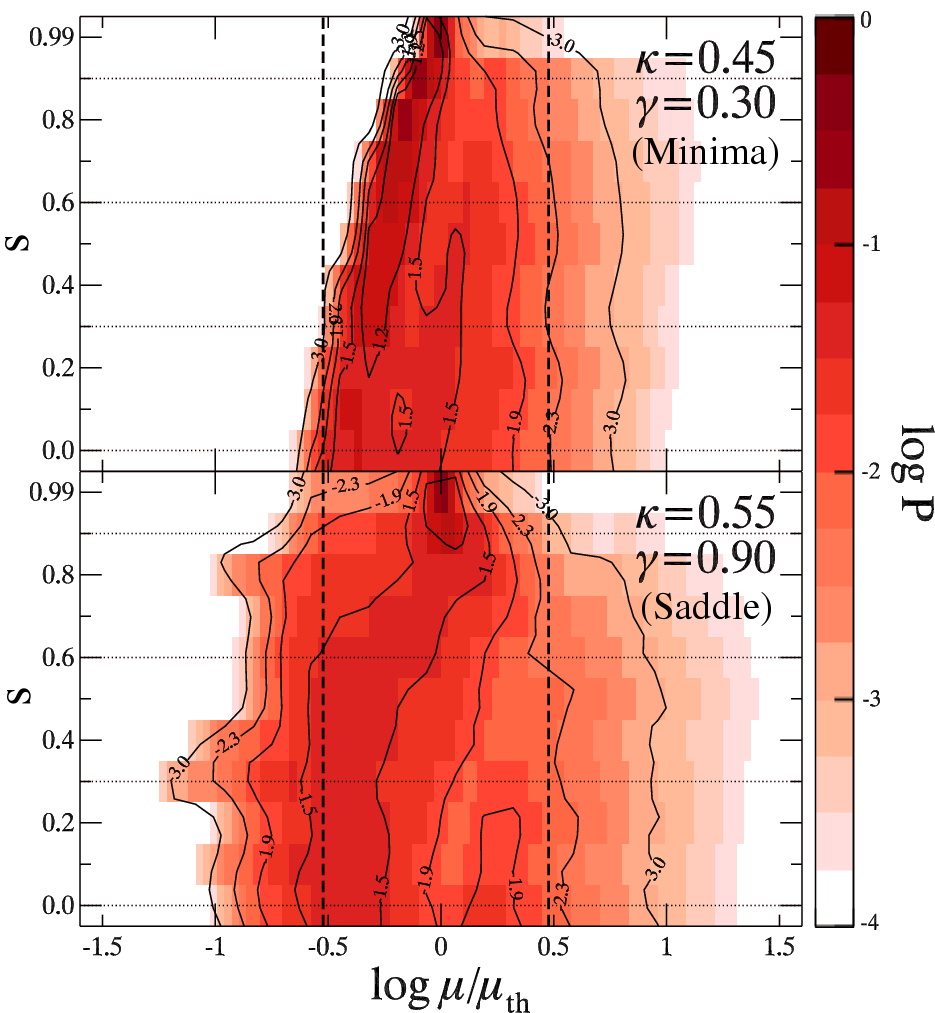}
\caption{Probability surfaces (shades of red online) with respect to the magnification and $s$. The magnification range is the same as in Figure \ref{fig:single}. Contours are drawn at ${\rm log  \, P} = (-1.25,-1.5,-1.9,-2.3,-3)$ to better depict the shape of the surfaces. The high resolution MPDs of Figure \ref{fig:single} can be seen here for the corresponding values of $s$ (horizontal dotted lines). The vertical dashed lines indicate $\mu=0.3\mu_{\rm th}$ and $\mu=3\mu_{\rm th}$. This Figure can be reproduced for any value of $\kappa,\gamma$ via the tools section of the GERLUMPH server.\label{fig:Psurface}}
\end{figure}

In Figure \ref{fig:single}, we also show for comparison the MPDs for 3 different smooth matter fractions, $s=0.3,0.6,0.9$.
It is expected that as the smooth matter fraction is varied, the shape of the MPD changes (more smooth matter means less microlenses).
In Figure \ref{fig:Psurface} we show the probability surface as a function of the magnification and smooth matter fraction. This representation highlights the changing width of the MPD, and the appearance of additional higher-probability peaks in the MPDs occuring between $0.3 \simeq \mu/\mu_{\rm th} \simeq 3.0$.
These additional peaks are related to the number of extra microimage pairs, which increase near the critical line \citep{Granot2003}.
We stress here that our online tools (Section 4) make these comparisons straightforward for any point in parameter space.

\subsection{Mass-sheet degeneracy}
\begin{figure}[t]
\plotone{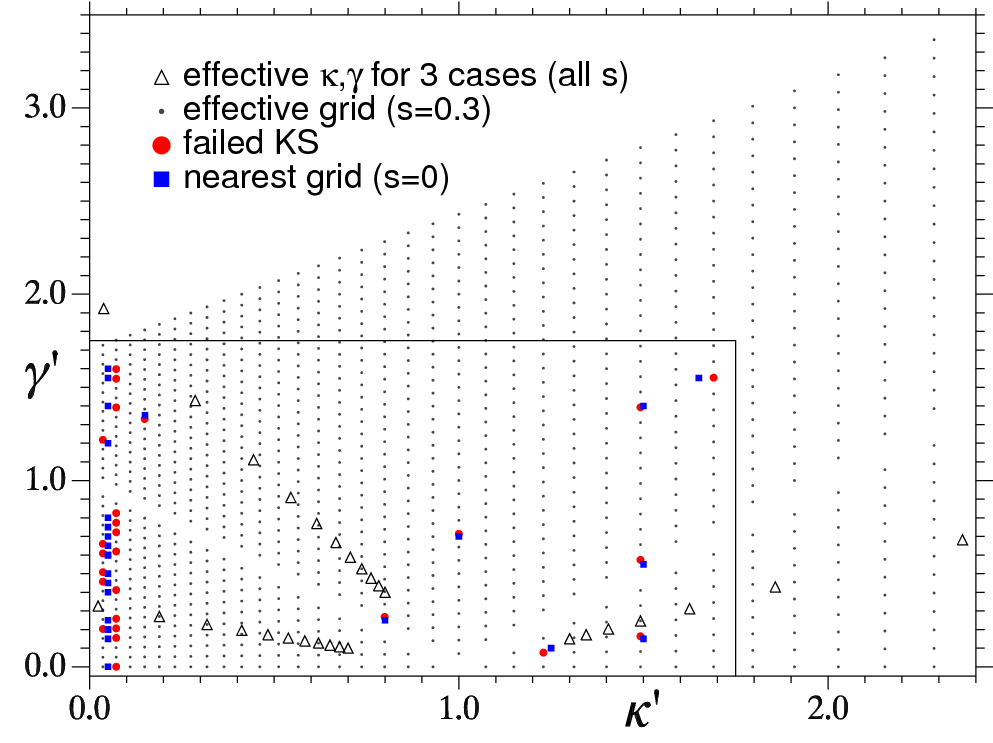}
\caption{The effective $\kappa',\gamma'$ parameter space. For three combinations of $\kappa,\gamma = (0.7,0.1), (0.8,0.4), (1.3,0.15)$, open triangles show the equivalent $\kappa',\gamma'$ for all values of $s$ (within the plot range) that we have used. The dots correspond to the effective grid of the subset of GD1 maps with $s=0.3$. Filled circles (red online) indicate the cases that failed the KS test with their nearest $s=0$ grid point, shown as a filled square (blue online). The rectangular region corresponds to the range shown in Figure \ref{fig:existing}.\label{fig:eff_pspace}}
\end{figure}

While we have chosen to perform a parameter survey by uniformly varying $\kappa$, $\gamma$ and $s$ (Figure \ref{fig:existing}), this is not the only possible sampling strategy.
The mass-sheet degeneracy is a transformation between equivalent lens models that produce the same observables (see Section \ref{sec:macro}).
In our case, the three-dimensional parameter space, $\kappa-\gamma-s$, transforms into the two-dimensional effective parameter space, $\kappa'-\gamma'$, through:
\begin{equation}
\label{eq:trans}
\kappa' = \frac{(1-s)\kappa}{1-s\kappa} \, , \, \gamma' = \frac{\gamma}{1-s\kappa}
\end{equation}
In Figure \ref{fig:eff_pspace}, we show the $\kappa',\gamma'$ locations for all the GD1 maps with $s=0.3$ (black dots).
The rectangular area of Figure \ref{fig:eff_pspace} corresponds to the $\kappa,\gamma$ range shown in Figure~\ref{fig:existing}, our original $\kappa,\gamma$ grid of 1122 points, and encloses 728 of the $\kappa',\gamma'$ pairs of the $s=0.3$ subset.
It can be seen from this example that the relevant GD1 maps do not cover the effective parameter space uniformly i.e. more densely for $\kappa' < 1$ (minima region) and more sparsely for $\kappa' > 1$.

To further illustrate the relationship between the GD1 maps and the effective parameter space, we present three $\kappa,\gamma$ combinations: (0.7,0.1), (0.8,0.4) and (1.3,0.15), one from each of the minima, saddle-point and maxima regions of parameter space.
For each case we calculate $\kappa',\gamma'$, shown in Figure~\ref{fig:eff_pspace} as open triangles, for all available steps in smooth matter fraction.
As the smooth matter content is increased, the equivalent $\kappa',\gamma'$ pairs move outwards radially from $\kappa',\gamma' = (1,0)$, as expected; that is the reason why for $\kappa,\gamma=$(1.3,0.15) maps with $s>0.7$ lie outside the range of the plot.

\begin{deluxetable*}{rrrrrc}
\tabletypesize{\scriptsize}
\tablewidth{0pt}
\tablecaption{KS test results for all values of $s$\label{tab:failed}}
\tablehead{\colhead{s} & \colhead{$N_{\rm total}$} & \colhead{$N_{\rm failed}$} & \colhead{$N_{\rm total}^{\kappa \geq 0.05}$} & \colhead{$N_{\rm failed}^{\kappa \geq 0.05}$} & \colhead{\% failed for $\kappa \geq 0.05$}}
\startdata
0.1    & 953 & 37  & 920 & 36 & 3.9 \\
0.2    & 822 & 24  & 789 & 22 & 2.8 \\
0.3    & 728 & 26  & 695 & 20 & 2.9 \\
0.4    & 633 & 27  & 600 & 18 & 3.0 \\
0.5    & 562 & 26  & 529 & 11 & 2.0 \\
0.6    & 438 & 27  & 420 &  6 & 1.4 \\
0.7    & 436 & 42  & 373 & 10 & 2.7 \\
0.8    & 423 & 60  & 269 &  4 & 1.5 \\
0.9    & 348 & 103 & 180 &  4 & 2.2 \\
0.99   & 317 & 276 &  10 &  0 & 0.0 \\
\enddata
\tablecomments{Number of total, $N_{\rm total}$, and failed, $N_{\rm failed}$, cases for the full effective grid, as well as for the grid with $\kappa \geq 0.05$, for all values of smooth matter fraction $s$. The failed cases for $\kappa \geq 0.05$ are also shown as a percentage over the total in the last column.}
\end{deluxetable*}

The mass-sheet degeneracy suggests that a given GD1 map should give the same microlensing outcomes as the effective-space map.
While the number of microlenses will almost be equal, the actual caustic networks on the equivalent maps will be different because we have used random microlens positions for all our maps.
However, their corresponding MPDs should be statistically equivalent.
Moreover, since a given $\kappa',\gamma'$ will not necessarily coincide with our original grid, we make our comparisons with the nearest grid point.

To examine the equivalence of the full parameter space with the effective space, we performed a Kolmogorov-Smirnov (KS) test between each of the 728 $\kappa',\gamma'$ maps with $s=0.3$ and the nearest $\kappa,\gamma,s=0$ map from GD1.
The null hypothesis was that the MPDs are the same.
Only 26 nearest neighbor pairs had a $p$-value less than 0.05, meaning that they failed the test, and they are shown as filled circles in Figure \ref{fig:eff_pspace}.
The positions of the $s=0$ GD1 maps with which these cases were compared are shown as filled squares.
It can be seen that the effective grid points lie quite far from our grid points with $\kappa = 0.05$, leading to a clustering of cases that failed the KS test for $\kappa < 0.1$.
This can be true for a few of the other failed cases as well, however, we point out that the latter ones fall in regions of parameter space where random microlens position systematics may play a role (see VF13, figure 6a).

We have performed a similar analysis for the remaining values of smooth matter fraction in GD1.
We expect more effective grid points to appear for very low values of $\kappa'$ as the smooth matter fraction is increased e.g. open triangles in the minima region of Figure \ref{fig:eff_pspace}.
The nearest original grid points to those cases will have $\kappa=0.05$, the lowest value in GD1, which may be too far away to have a similar MPD and consequently fail the KS test.
This can be seen in the first three columns of Table \ref{tab:failed}, which show the total number of effective grid points lying inside our original grid and the number of cases that failed the KS test, as $s$ is increased.
For very high $s$, the majority of effective grid points lie below $\kappa=0.05$ and so have a statistically different MPD, e.g. for $s=0.99$ all the failed cases lie below $\kappa=0.05$, where also 97\% of the total effective grid points are located.
However, if we take into account only the effective grid points with $\kappa \geq 0.05$, the number of cases that failed the KS test lies between 0 and 4\% (three last columns of Table \ref{tab:failed}).
These cases can be effected by random microlens position systematics, or it can be again that the nearest grid point lies too far to have a similar MPD.

The above experiments show that the MPDs are similar throughout the examined parameter space, and are in agreement with what is expected from the mass-sheet degeneracy.

\subsection{Magnification fluctuations}
\label{sec:fluct}
As introduced in Section \ref{sec:fluct_intro}, microlensing induced variations towards high and low magnifications are expected to affect different kinds of observations.
High magnification events \citep[e.g.][]{Anguita2008} are effected by the high magnification tail of the MPDs, while anomalous flux systems more likely occur due to the demagnification of one of the images due to a combination of both smooth matter and compact objects \citep{Schechter2002}.
In order to investigate how smooth matter effects the high magnification part of the MPDs, we define a quantity, $P_{3}$, that has been used in the past \citep{Rauch1992,Wambsganss1992} to measure the total probability in the magnification range $3 \mu_{\rm th} < \mu < +\infty$ (right of the dashed line in Figure \ref{fig:Psurface}):
\begin{equation}
P_{3} = \int_3^{+\infty} P(\mu/\mu_{\rm th}) d(\mu/\mu_{\rm th}) = \sum^{\mu_i>3\mu_{\rm th}} P(\mu_i) ,
\end{equation}
A similar quantity is defined for demagnifications:
\begin{equation}
P_{0.3} = \int_{-\infty}^{0.3} P(\mu/\mu_{\rm th}) d(\mu/\mu_{\rm th}) = \sum^{\mu_i<0.3\mu_{\rm th}} P(\mu_i) ,
\end{equation}
which is the total probability in the magnification range $-\infty < \mu < 0.3 \mu_{\rm th}$ (left of the dashed line in Figure \ref{fig:Psurface}).
We examine the effect of smooth matter on these two quantities throughout the original $\kappa,\gamma,s$ parameter space.

\begin{figure}[t]
\plotone{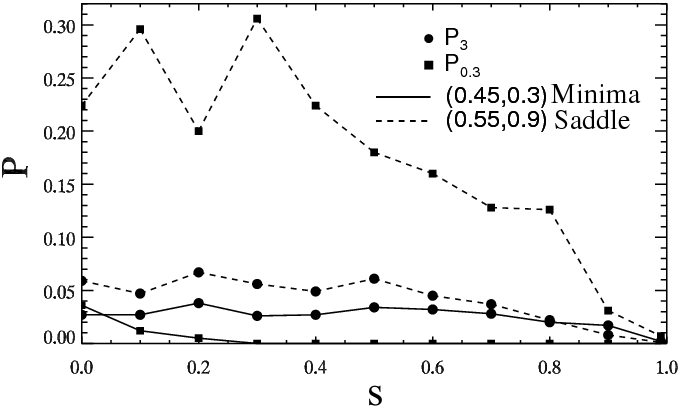}
\caption{Value of $P_3$ (circles) and $P_{0.3}$ (squares) as a function of $s$, for $\kappa,\gamma$ equal to $(0.45,0.3)$ (dashed line) and $(0.55,0.9)$ (solid line). This Figure can be reproduced for any value of $\kappa,\gamma$ via the tools section of the GERLUMPH server.\label{fig:single_int}}
\end{figure}

Figure \ref{fig:single_int} shows the variation of $P_{3}$ (filled circles) and $P_{0.3}$ (filled squares) as $s$ is increased, for two trial combinations $\kappa,\gamma = (0.45,0.3)$ (solid lines) and $\kappa,\gamma = (0.55,0.9)$ (dashed lines).
The situation is quite different between the minima and saddle-point regions of parameter space.
It is clear that the saddle-point case is much more demagnified than the minima one.
Moreover, increasing the presence of smooth matter past $s=0.2$ supresses demagnifications in the minima region.
High magnifications for the saddle-point case are two times more likely than the minima case.
Both high and low magnifications start to decrease past roughly $s=0.5$ for the saddle-point case, while the decrease is slower and starts around $s=0.6$ for high magnifications for the minima.

\begin{figure}[t]
\plotone{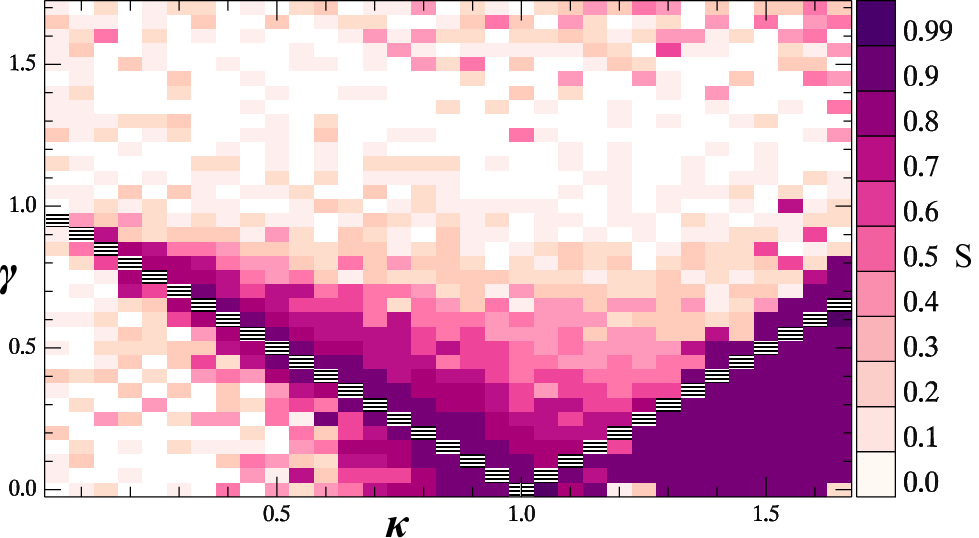}
\caption{Value of $s$ for which the maximum probability for magnifications higher than $3 \mu_{\rm th}$ occurs, shown across the parameter space. The presence of smooth matter enhances the high magnification fluctuations in the maxima region and the region along the critical line. For $s=0.99$, the probability drops, thus, setting an upper limit to this enhancement.\label{fig:max_high}}
\end{figure}

\begin{figure}[t]
\plotone{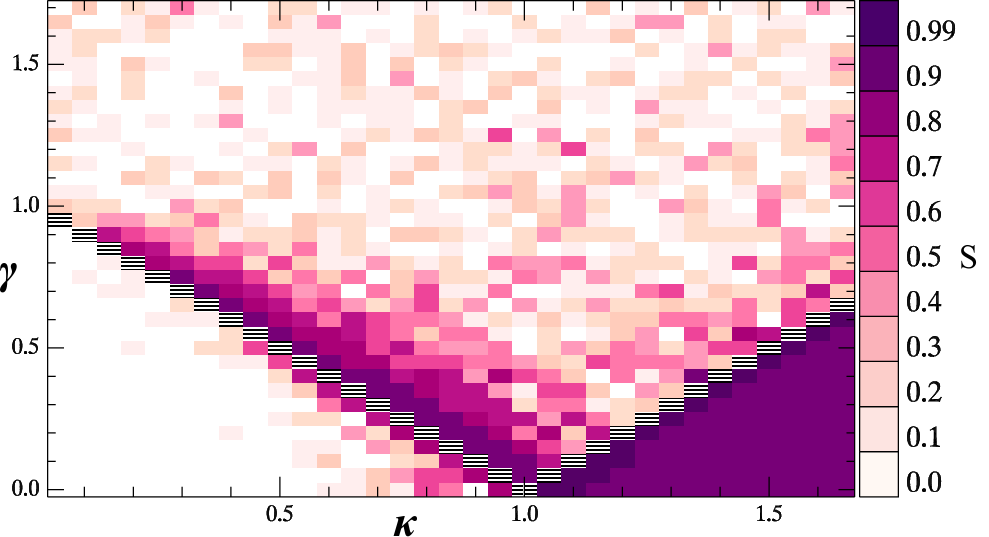}
\caption{Value of $s$ for which the maximum probability for magnifications lower than $0.3 \mu_{\rm th}$ occurs, shown across the parameter space. The presence of smooth matter enhances the high demagnification fluctuations in the maxima region and the region just above the critical line for $\kappa < 1$.\label{fig:max_low}}
\end{figure}

By performing a similar analysis in other $\kappa,\gamma$ locations, we can identify the general trend of $P_3$ and $P_{0.3}$ with respect to $s$ across parameter space regions.
In the minima and saddle-point regions, there is a value of $s$ which gives the maximum probability for high/low magnifications.
This can be seen in Figures \ref{fig:max_high} and \ref{fig:max_low}, where we show the value of $s$ for which the maximum $P_3$ and $P_{0.3}$ occur, for a given $\kappa,\gamma$ combination.
The maximum never occurs at $s=0.99$ because there are almost no compact objects, which confirms our expectation that there is a limit to the enhancement of microlensing induced fluctuations due to smooth matter \citep{Schechter2002}.

\begin{figure}[t]
\plotone{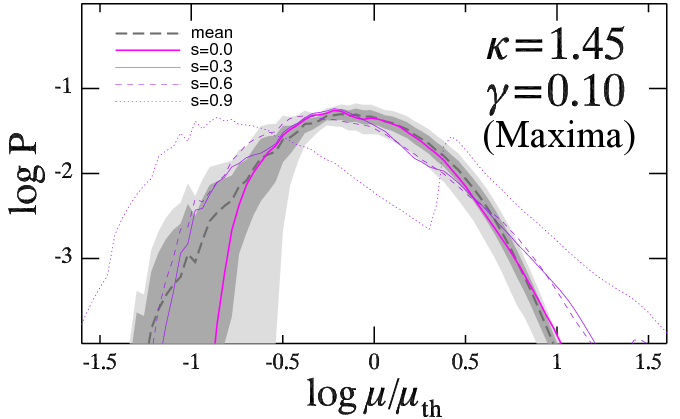}
\caption{Example magnification probability distribution (MPD) from the maxima region, for $\kappa,\gamma = (1.45,0.1)$ (similar to Figure \ref{fig:single}). The thin solid, dashed and dotted lines (purple online) are the MPDs for $s$ equal to 0.3, 0.6 and 0.9 respectively. The probability for very high magnifications and demagnifications is maximized for $s=0.9$. The mean MPD for $s=0$, thick dashed line, and the one standard deviation region, grey area, from VF13 are also shown for comparison.\label{fig:maxima}}
\end{figure}

The maxima region is different than the minima and saddle regions, since the maximum $P_3$ and $P_{0.3}$ occur consistently at $s=0.9$ throughout it.
With closer examination, there are high oscillations for low values of $s$, which shift the maximum of $P_{3}$ to lower $s$ in three cases: $\kappa = (1.2, 1.25, 1.65)$ and $\gamma = 0.0$, but the overall behaviour does not change.
By identifying the particular behaviour of the maxima region from Figures~\ref{fig:max_high} and \ref{fig:max_low}, we can now go back and examine the MPDs in detail for an example case.
In Figure~\ref{fig:maxima}, we show the MPDs for 4 different smooth matter fractions, $s=0.0,0.3,0.6,0.9$, for a trial combination of $\kappa,\gamma = (1.45,0.1)$ in the maxima region (the mean MPD from VF13 for $s=0$ is also shown).
The MPD for $s=0.9$ (dotted line) is the one that suppresses magnifications between $0.3 < \mu/\mu_{\rm th} < 3.0$ and provides the most enhanced fluctuations from $\mu_{\rm th}$.

\section{The web server and online tools}
\label{sec:server}
To support online access to the GERLUMPH maps, we have developed a series of tools that can be used in a web browser with JavaScript support enabled.
The GERLUMPH server is based on the open-source LAMP software bundle \citep[Linux/GNU Apache MySQL PHP,][]{Lee2002}.
For specific applications we find that WebGL\footnote{Web Graphics Library is a JavaScript API for rendering interactive 3D and 2D graphics within any compatible web browser, allowing for GPU acceleration.} provides a performance boost, especially for online image processing.
In the following, we outline the data and services provided by the server.
A more detailed guide with examples of how to use the GERLUMPH resource can be found online\footnote{http://gerlumph.swin.edu.au/guide/}.

The GERLUMPH maps can be retrieved either by a direct query to the map database\footnote{http://gerlumph.swin.edu.au/parsquery/} or by selecting a specific system and a corresponding macro-model\footnote{http://gerlumph.swin.edu.au/macro-models/}.
A user can then proceed with downloading the maps, or use one of the provided tools to inspect the map properties.

Currently, we provide online tools to generate graphs similar to Figures \ref{fig:single}, \ref{fig:Psurface} and \ref{fig:single_int}, but for any combination of $\kappa,\gamma$, from GD1.
There are also tools for reproducing the GD0 data visualizations presented in VF13.
Finally, we introduce the `Colorbar', a tool that color-highlights caustics of different magnification on a map, together with the corresponding part of the MPD.
An example of the use of `Colorbar' can be seen in Figure \ref{fig:colorbar}.
All the above tools, together with detailed descriptions and examples, can be accessed at the results section of the GERLUMPH website\footnote{http://gerlumph.swin.edu.au/\#tools}.

We envisage several modes of operation:
\begin{enumerate}
\item Particular microlensed systems of interest can be studied by accessing individual parameter-space values, typically through the macro-model tool.
\item For use in external, pre-existing analysis tools or to compare magnification maps or MPDs produced by other means, a small subset of maps would be selected for downloading.
\item For specific parameter-space explorations requiring from 100s to 1000s of GD1 maps, new e-tools should be implemented and made accessible through the web service. This preferred paradigm of pushing the computation to the data \citep{Szalay2001,Gray2004} obviates the need for individuals to download (potentially) terabytes of data for local analysis.
\end{enumerate}

\section{Discussion}
We have presented a first application of the GD1 results: a detailed numerical investigation of the mass-sheet degeneracy.
Using the transformations to effective convergence and shear (\ref{eq:trans}), we have compared maps with a variety of smooth matter values to those with $s=0$, over a region of parameter space relevant to microlensing research ($0.0<\kappa,\gamma<1.7$, Figure \ref{fig:eff_pspace}).
Fewer than 4\% of cases failed the KS test for any value of $s$ available in the GD1 maps (Table \ref{tab:failed}).
The reasons why these cases fail the KS test are well understood: either the nearest original GD1 parameter point was too far from the effective one and thus had a significantly different MPD, or the maps were in a region of parameter space where microlens position systematics should have been taken into account (VF13).


The transition in the MPD shapes in Section \ref{sec:results} (Figures \ref{fig:single} and \ref{fig:Psurface}) can be understood in terms of the analytic result of \cite{Granot2003}, who attribute the appearence of each peak to the presence of an extra pair of microimages.
Increasing $s$ causes the MPDs to become narrower and reduces the probability for extra microimage pairs.
While \cite{Granot2003} compared their analytic result only to 3 cases due to computational limitations, GERLUMPH has enabled such tests anywhere in parameter space.
For the moment, such a test cannot be performed online, but users are welcome to download any map and use it with their own tools.

Another application of GD1 is exploring the effect of smooth matter on the high magnification and demagnification parts of the MPDs in the $\kappa,\gamma$ parameter space.
Including smooth matter increases the total probability of high magnification along the critical line and for increasing $\kappa$ (Figure \ref{fig:max_high}).
In the rest of the parameter space, away from the critical line and the maxima region, the probability of high magnifications is decreased.
For demagnifications, the minima, saddle-point and maxima regions show distinct properties.
Smooth matter has almost no effect in the minima region, while just above the critical line, in the saddle-point region, high values of smooth matter ($s=0.7-0.9$) enhance the low magnification part of the MPD.
This effect is slowly reduced as we get further from the critical line.
Finally, $s=0.9$ provides the maximum demagnification probability throughout the maxima region.

Our experiments in Section \ref{sec:fluct} allow us to confirm the findings of \cite{Schechter2002}, who considered the close pair of images of the lensed quasar MG0414+0534, with $\kappa,\gamma = (0.475,0.425)$ (minimum) and $\kappa,\gamma = (0.525,0.575)$ (saddle), and found an enhancement in demagnification when including smooth matter.
However, our results extend this picture to the entire parameter space, indicating that this only happens close to the critical line, and not in the rest of minima or saddle-point regions.

\cite{Vernardos2013} found that there are specific areas of parameter space where the MPD is effected by the size of the map, and more than one map, or a wider map, would be needed to get a representative behaviour.
Our present results (Figures \ref{fig:max_high} and \ref{fig:max_low}) appear consistent throughout these areas, meaning that the uncertainty in the low magnifications does not play an important role in the properties that we are examining.

Finally, consistently throughout the maxima region we find that the strongest microlensing induced fluctuations occur for $s = 90\%$.
This result, together with the results of VF13, indicate that different microlensing signatures should be expected in the maxima region of parameter space.
However, maxima images of background quasars are expected to occur near the central regions of the galaxy-lens, where the smooth matter fraction should be quite low.

\section{Future: eResearch and the microlensing cloud}
As the volume of astronomical data continues to increase, there is a growing need to move more analysis tasks to a remote service model on cloud-like architectures \citep[e.g.][]{Berriman2011}.
This approach of pushing the research question to the data, rather than the traditional method of bringing the data to the desktop, sits at the core of the  Virtual Observatory (VO) philosophy \citep{Szalay2001,Gray2004}.

While most observational datasets are now routinely published in VO-compliant online archives, VO-theory is lagging behind \citep[see][for an early discussion]{Teuben2002}.
In part, this is because it has proven complex to define an appropriate data standard, and so each field has focused on its own needs and implementations, such as astrophysical computational simulations \citep{Cassisi2008,Lemson2012}.

With GD1, we provide an in-progress case study on how to manage a growing on-line data archive, where the preferred analysis method is to use in-browser tools.
As we continue to generate and add additional maps to the GERLUMPH archive, we anticipate that the final data set will comprise $\sim 60000$ high resolution maps, corresponding to $\sim 20$ TB of data.
The number of maps and the amount of data is not too large to be reprocessed, but it is not small enough that it can be easily downloaded and used by individuals.
Indeed, it would be very time and resource consuming to download all, or even significant subsets of these maps, in order to perform parameter space studies of cosmological microlensing.

We are continuing to investigate ways to implement and provide additional tools for online processing, including supporting map and source profile convolutions, automated and 
interactive extraction of light curves, and an external map upload facility to facilitate model and technique-based comparisons.

In combination with observational databases (e.g. CASTLES\footnote{http://www.cfa.harvard.edu/castles/} or The Master Lens database\footnote{http://masterlens.astro.utah.edu/}) and other online applications \citep[e.g. Mowgli]{Naudus2010}, GERLUMPH will become a major piece of a new microlensing eResearch cloud, effectively redesigning the way we approach the analysis and investigation of quasar accretion discs and supermassive black holes in the synoptic survey era.

\section{Conclusions}
\label{sec:conclusions}
As new synoptic all-sky surveys commence, the number of known gravitationally lensed quasars is set to increase by a factor of ten, from around one hundred \citep{Mosquera2011b} to, potentially, a few thousand \citep{Oguri2010}.
These discoveries will enable improved understanding of the role that quasars and their central supermassive black holes play in the formation and evolution of galaxies and the large scale structure of the Universe.
Obtaining the best lensing-based constraints on quasar/black hole properties requires extensive computational modeling, through the generation of microlensing magnification maps.
There is a need now to fully explore the connection between inferred quasar model constraints (such as the size of emission regions, or the accretion disc temperature profile) and the underlying microlensing model.

GERLUMPH is a cosmological microlensing theoretical parameter survey that takes up the challenge of parameter space exploration.
GERLUMPH magnification maps can be used with standard analysis techniques (e.g. convolution with source profiles and/or light curve extraction) for applications to observations.
However, it is not our intention that GERLUMPH should remove the need for individuals to produce their own magnification maps: indeed such computations and comparisons to GERLUMPH datasets would be encouraged.
What makes GERLUMPH unique is the ability to carry out studies systematically across the microlensing parameter space, using a set of online e-tools.
This toolbox can be enriched at any time by implementing additional analysis or visualisation strategies the microlensing community may have to suggest.

We have used the 12342 maps of GD1 to test the mass-sheet degeneracy, and our results are in agreement with its predictions.
We have explored the effect of smooth matter on high magnifications ($>3 \mu_{\rm th}$) and demagnifications ($<0.3 \mu_{\rm th}$), and found that the presence of smooth matter enhances high magnifications along the critical line and in the maxima region, while demagnifications are enhanced mostly in the saddle-point and maxima regions.
Our GD1 data (together with the GD0 dataset used in VF13) are publicly available and can be downloaded through our web server, which also provides an initial set of online analysis tools.

We have gained new insight into the management of moderate-sized ($\sim20$ TB) theory-based astronomy datasets, which will be of benefit to future eResearch projects in astronomy and other sciences.
The GERLUMPH data resource aims to provide a computationally demanding piece of the web-based microlensing computing cloud, and enables a new approach for generating high quality results to interpret upcoming new discoveries of gravitationally lensed quasars.

\acknowledgements
This research was undertaken with the assistance of resources provided at gSTAR through the ASTAC scheme supported by the Australian Government.
gSTAR is funded by Swinburne and the Australian Government's Education Investment Fund.
NFB thanks the Australian Research Council (ARC) for support through Discovery Project (DP110100678).
DC acknowledges support through an ARC QEII Fellowship.
We thank the anonymous referee for useful comments on this work.

\bibliographystyle{mn2e}
\bibliography{gerlumph}

\end{document}